\documentclass{ws-procs975x65}

\newcommand{\chushi}[1]{}

\begin{document}

\title{125 GeV Technidilaton at the LHC
}

\author{Shinya Matsuzaki~\footnote{
S.M. is currently at Institute for Advanced Research, Nagoya University, Nagoya 464-8602, Japan. 
}}

\address{Maskawa Institute for Science and Culture, Kyoto Sangyo University, Motoyama, Kamigamo, Kita-Ku, Kyoto 603-8555, Japan.}



\begin{abstract}
The technidilaton (TD) is a composite scalar  
predicted in walking technicolor (WTC), arising as a pseudo Nambu-Goldstone boson 
associated with the spontaneous breaking of the approximate scale invariance. 
Due to the Nambu-Goldstone boson's nature, 
the TD can be as light as the LHC boson that has been discovered at around 125 GeV. 
We discuss the size of the TD mass and the coupling properties relevant to 
the LHC study. 
It turns out that the TD couplings to the standard model (SM) particles take the same form 
as those of the SM Higgs boson, except the essentially distinguishable two ingredients: 
i) the overall coupling strengths set by the decay constant related to the spontaneous breaking 
of the scale invariance, which is in general not equal to the electroweak scale; 
ii) the couplings to photons and gluons which can include extra contributions 
from technifermion loops and hence can be enhanced compared to the SM Higgs case. 
To be concrete, we take the one-family technicolor model to explore the TD 
LHC phenomenology at 125 GeV. 
It is shown that the TD gives the signal consistent with the currently reported 
LHC data, notably can explain the excess in the diphoton channel, 
due to the extra contributions to digluon and diphoton couplings coming from 
the one-family technifermion loops.

\end{abstract}

\bodymatter

\section{Introduction}

On July 4, 2012, a new particle at around 125 GeV was discovered at the LHC~\cite{Aad:2012tfa}. 
Through measurements of the coupling property, spin and parity, 
the new particle has so far been almost consistent with the Higgs boson predicted 
as a key boson responsible for the origin of mass in the standard model (SM). 
One discrepancy from the SM Higgs has been, however, reported in the diphoton channel~\cite{ATLAS-CONF-2012-168} 
where the observed signal event is about two times larger than the SM Higgs prediction. 
This would imply that the observed scalar boson is a SM-Higgs impostor concerning 
the underlying theory beyond the SM.

One such a possible impostor is the technidilaton (TD), a composite scalar boson, predicted in 
the walking technicolor (WTC)~\cite{Yamawaki:1985zg,Bando:1986bg} 
with an approximately scale-invariant (conformal) 
 gauge dynamics and  a large anomalous dimension
$\gamma_m =1$.  
The TD  is a pseudo Nambu-Goldstone boson for the spontaneous breaking of 
the approximate scale symmetry triggered by technifermion condensation 
and hence its lightness, say 125 GeV,  
would be protected  by the approximate scale symmetry inherent to the WTC.
Thus the discovery of TD would imply the discovery of the WTC.

The LHC signatures of the TD were studied in a couple of recent papers~\cite{Matsuzaki:2011ie,Matsuzaki:2012vc,Matsuzaki:2012xx}.  
It was shown~\cite{Matsuzaki:2012vc} that 
the 125 GeV TD is consistent with the currently reported diphoton signal as well as 
other signals such as $WW^*$ and $ZZ^*$, etc.. 
 It was emphasized that the TD is favored by the current data 
thanks to the presence of extra technifermion loop corrections to digluon and diphoton couplings.

This talk summarizes the 125 GeV TD phenomenology at the LHC and shows 
a theoretically interesting possibility that the TD can indeed be as light as 
the 125 GeV boson by certain nontrivial feature intrinsic to the dynamics of WTC.  
The stability of the TD mass is also discussed, where it turns out that 
the large TD decay constant plays an important role against 
corrections from the SM particles in a low-energy region 
to be relevant after the walking behavior.

We start with a review of the characteristic features of WTC including the spontaneous breaking of 
both chiral and scale symmetries and the mass generation of TD (Sec.~\ref{WTC-TD}). 
We then discuss the size of the TD mass following two approaches: 
straightforward nonperturbative computations in the ladder approximation~\cite{Harada:2003dc,Harada:2005ru,Shuto:1989te} 
(Sec.~\ref{lad-est}) 
and a model estimate in a view of holography applied to WTC~\cite{Haba:2010hu,Matsuzaki:2012xx} (Sec.\ref{holo-est}). 
In contrast to the ladder estimate, the holographic model incorporates effects from the technigluon condensation,  
which turns out to play the crucial role to realize the TD mass on the order of 125 GeV~\cite{Matsuzaki:2012xx}.

To make direct contact with the LHC study, 
we derive a low-energy effective Lagrangian~\cite{Matsuzaki:2012vc} 
obtained by integrating out technifermions, 
which is governed by the lightest technihadron spectra involving the TD (Sec.~\ref{eff-lag}). 
The Lagrangian is constructed based on nonlinear realization of 
the scale symmetry as well as the chiral symmetry coupled to the SM particles via 
the SM gauge interactions.

The TD couplings to the SM particles are read off from the effective Lagrangian.  
We then find that the couplings take essentially the same form as those of the SM particles, 
except two distinguishable ingredients: the overall scale of the couplings 
are set by the TD decay constant ($F_\phi$) associated with the spontaneous breaking of the 
scale symmetry, not by the electroweak (EW) scale ($v_{\rm EW}$). 
Actually, $F_\phi$ turns out to be larger than $v_{\rm EW}$ from both ladder and holographic estimates. 
The other crucial discrepancy comes in the couplings to diphoton and digluon, which get effects from 
technifermion loop contributions to be either enhanced or suppressed, depending on models of WTC. 
To be concrete, we will take one-family model for the WTC to explicitly calculate those couplings. 
After understanding the TD coupling property from the effective Lagrangian, 
we pay attention to the stability of the TD mass against radiative corrections arising from 
the SM particle loops. 
The large TD decay constant $F_\phi$ then turns out to play the crucial role 
to keep the TD mass around 125 GeV without invoking heavy fine tunning, in contrast to 
the SM Higgs case.

Finally, we shall explore the LHC production and decay processes of the 125 GeV TD in the one-family model  
and estimate the signal strengths for the relevant channels to be compared with 
the experimental data on the Higgs searches (Sec.~\ref{signal}). 
It is shown that the TD signal is consistent with the currently reported data and 
can give a better fit than the SM Higgs, notably due to the presence of 
extra technifermion loop corrections.

\section{Walking Technicolor and Technidilaton}\label{WTC-TD}

Technicolor (TC)~\cite{Weinberg:1975gm,Farhi:1980xs} accommodates  
the EW symmetry breaking by technifermion condensation, without invoking the fundamental Higgs boson,  
just like the quark condensation in QCD, and hence gives the dynamical explanation for the origin of mass. 
The original version of TC~\cite{Weinberg:1975gm}, 
a naive scale-up version of QCD,  was ruled out due to the excessive 
flavor-changing neutral currents.

A way out of this problem was suggested under a simple assumption of the existence of 
 a large anomalous dimension for technifermion bilinear operator $\gamma_m$ 
without any concrete dynamics and concrete  
 value of the anomalous dimension~\cite{Holdom:1981rm}.  
It  was the WTC~\cite{Yamawaki:1985zg,Bando:1986bg} that exhibited a concrete dynamics based on 
a nonperturbative analysis of ladder Schwinger-Dyson (SD) equation with
nonrunning (scale invariant/conformal) gauge coupling, $\alpha(p) \equiv \alpha$, 
yielding a concrete value of the anomalous dimension, $\gamma_m = 1$ 
in the chiral broken phase.  
Modern version of WTC~\cite{Lane:1991qh,Appelquist:1996dq, Miransky:1996pd} is based on the two-loop running coupling 
with the Caswell-Banks-Zaks infrared fixed point~\cite{Caswell:1974gg},
 instead of the nonrunning one, in the improved ladder SD equation.

Another problem of the TC as a QCD scale-up is the EW constraints, so-called $S$ and $T$  parameters. 
This may also be improved in the WTC~\cite{Appelquist:1991is,Harada:2005ru}.  
Even if  WTC in isolation cannot overcome this problem, there still exist a possibility that the problem may be 
resolved in the combined dynamical system including the SM fermion mass generation such as the extended TC 
(ETC) dynamics~\cite{Dimopoulos:1979es}, 
in much the same way as the solution (``ideal fermion delocalization'')~\cite{Cacciapaglia:2004rb} in the Higgsless models which simultaneously
adjust $S$ and $T$ parameters by incorporating the SM fermion mass profile.

\begin{figure}[t]  
\includegraphics[width=6.0cm]{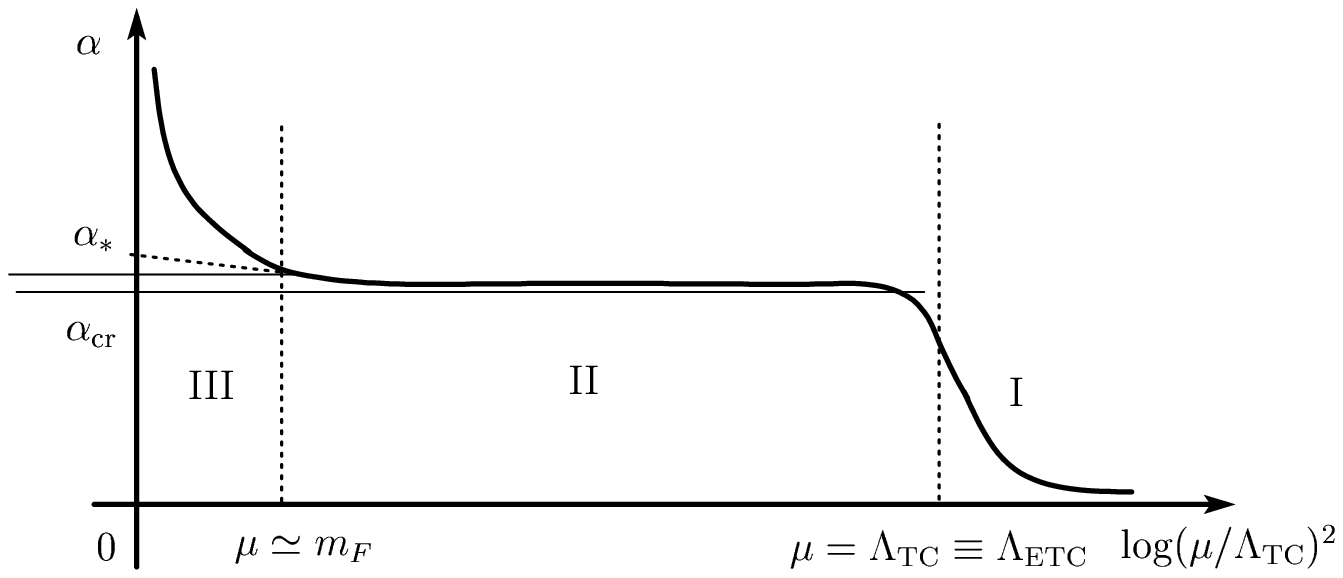}
  \includegraphics[width=6.0cm]{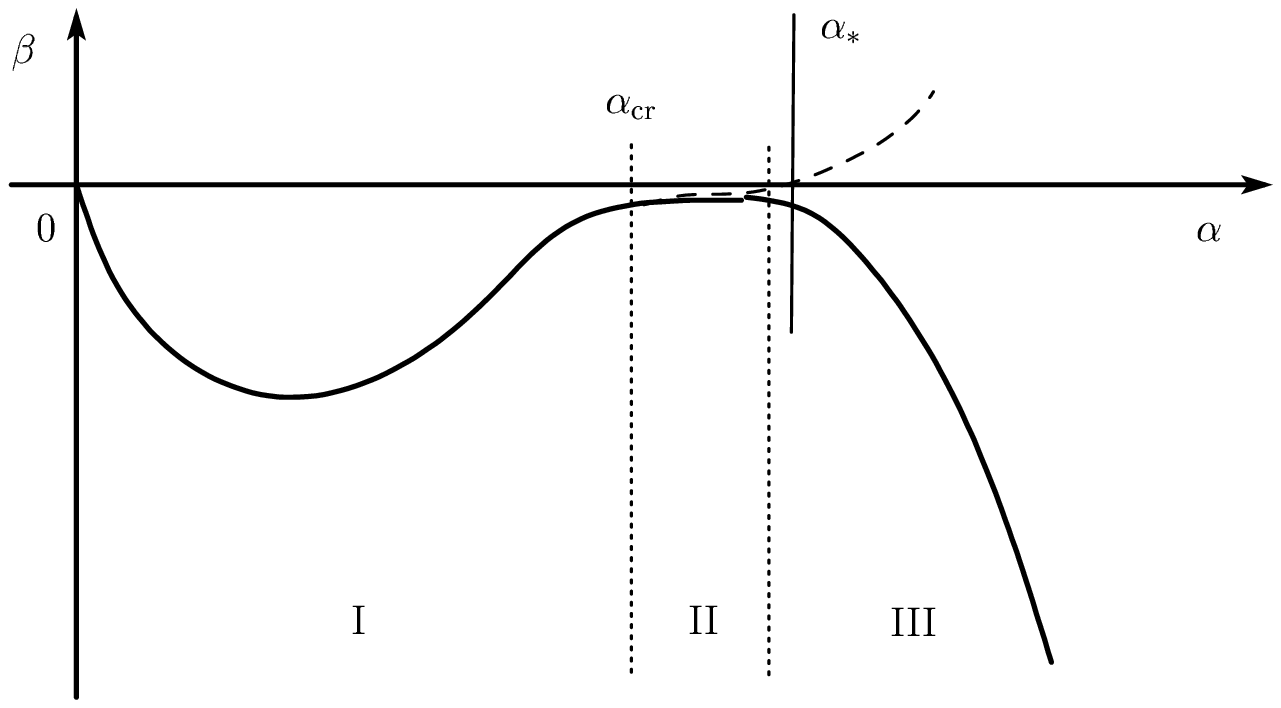}
\caption{A schematic view of WTC in terms of the running of the gauge coupling $\alpha$ (left panel) and the beta function $\beta$ (right panel). } 
\label{alpha-beta}
\end{figure}

It would be better to understand the WTC through an illustration by the notion of gauge dynamics: 
Figure \ref{alpha-beta} shows a schematic view of the WTC  
in terms of the gauge coupling $\alpha$ (left panel) and its beta function $\beta(\alpha)$ (right panel). 
In a field theoretical sense, the ``walking" can be realized by an accidental balance between 
gluonic and fermionic contributions to the running of gauge coupling of $SU(N_{\rm TC})$ gauge theory 
including $N_{\rm TF}$ fermions. 
In fact, the presence of walking region implies a pseudo infrared fixed point ($\alpha_*$), 
a la Caswell-Banks-Zack infrared fixed point based on the two-loop beta function in the large $N_f$ QCD~\cite{Caswell:1974gg}. 
During the walking region (region II in Fig.~\ref{alpha-beta}), 
the gauge coupling slowly reaches the critical coupling, which is slightly off from an infrared fixed point $\alpha_*$, 
$\alpha_{\rm cr}(< \alpha_*)$, where the chiral symmetry is dynamically broken by technifermion condensation $\langle \bar{F}F \rangle \neq 0$ 
and hence technifermions get the dynamical mass $m_F = {\cal O}(4 \pi F_\pi)\sim 1$ TeV, where 
$F_\pi$ denotes the technipion decay constant associated with the chiral symmetry breaking.  
The walking regime (region II) expands in a wide range $(m_F< \mu < \Lambda_{\rm TC})$: 
below the infrared scale $m_F(\sim 1 \, {\rm TeV})$, technifermions decouple and hence the balance with technigluon 
contributions gets lost, leading to the one edge of walking (region III), 
while above the ultraviolet scale $\Lambda_{\rm TC}(\sim 10^3-10^4$ TeV) the theory 
will be embedded into an ETC (region I).

The dynamical mass generation of technifermion is characterized by so-called Miransky scaling~\cite{Miransky:1984ef}, 
tied with the conformal phase transition~\cite{Miransky:1996pd}, 
\begin{equation} 
 m_F \sim \Lambda_{\rm TC} e^{-\frac{\pi}{\sqrt{\alpha/\alpha_{\rm cr} -1}}}
 \,, \qquad 
{\rm for} \qquad \alpha > \alpha_{\rm cr}
 \,. \label{Miransky}
\end{equation} 
Note that this scaling property implies the wide-range walking regime above $\alpha_{\rm cr}$ in region II of Fig.~\ref{alpha-beta}. 
The chiral condensate $\langle \bar{F}F \rangle$ is generated to scale with the large anomalous dimension 
$\gamma_m \simeq 1$ which solves the flavor-changing neutral current problem~\cite{Yamawaki:1985zg}: 
\begin{equation} 
 \langle \bar{F}F \rangle_{\Lambda_{\rm TC}} \sim \frac{N_{\rm TC}}{4\pi^2} m_F \Lambda_{\rm TC}^2 
 \,, \qquad 
  \langle \bar{F}F \rangle_{\Lambda_{\rm TC}}
  \simeq 
  \left( \frac{\Lambda_{\rm TC}}{m_F} \right)^{\gamma_m \simeq 1}
  \langle \bar{F}F \rangle_{m_F}
 \,. 
\end{equation}

Of particular interest is the fact that 
the WTC possesses the (approximate) scale invariance ($\beta(\alpha) \simeq 0$ for $m_F < \mu < \Lambda_{\rm TC}$),   
which is spontaneously broken by the technifermion condensation/mass generation. 
This implies presence of a (pseudo) Nambu-Goldstone boson (``dilaton") for the scale symmetry. 
This is a flavor-singlet composite scalar and may be a mixture of $\bar{F}F$ bound state and techniglueball. 
That is the TD, technidilaton.

The TD gets massive essentially due to the ``nonperturbative" scale anomaly: 
 According to the dynamical mass generation in Eq.(\ref{Miransky}), 
 the gauge coupling $\alpha$ is renormalized, starts running and hence  
the ``nonperturbative" beta function $\beta_{\rm NP}(\alpha)$ is generated~\cite{Bardeen:1985sm}: 
\begin{equation} 
\beta_{\rm NP}(\alpha) 
= \Lambda_{\rm TC} \frac{\partial \alpha}{\partial \Lambda_{\rm TC}}
= - \frac{2\alpha_{\rm cr}}{\pi} \left(\frac{\alpha}{\alpha_{\rm cr}}-1 \right)^{3/2}
\, \label{NP:beta}.  
\end{equation} 
 This induces the ``nonperturbative" scale anomaly: 
 \begin{equation}
  \partial_\mu D^\mu = \frac{\beta_{\rm NP}(\alpha)}{4 \alpha^2} \left( 
\alpha G_{\mu\nu}^2 \right)  \neq 0 
\,,  \label{NP:SA}
 \end{equation}
where $D_\mu$ denotes the dilatation current and $G_{\mu\nu}$ the field strength of technigluon field. 
Thus the TD becomes massive due to the nonperturbative scale anomaly~\cite{Yamawaki:1985zg,Bando:1986bg}.

\subsection{Technidilaton Mass: Ladder Estimate}\label{lad-est}

A straightforward calculation ~\cite{Harada:2003dc}
based on the  ladder SD equation and the ladder 
(homogeneous) Bethe-Salpeter (BS) equation in the walking regime indicates  a relatively light scalar bound state 
(identified with TD): 
\begin{equation} 
{\rm Ladder}: \qquad 
M_\phi \sim 4 F_\pi \simeq 500 \,{\rm GeV}
\,,   \label{mtd:lad}
\end{equation} 
for the one-family model with $F_\pi \simeq 123$ GeV. 
This $M_\phi$ is much smaller than masses of the techni-vector/axial-vector mesons on TeV range, 
but still larger than the LHC boson at 125 GeV. 
This result is consistent with another calculation~\cite{Harada:2005ru} based on 
the ladder SD equation and the ladder (inhomogeneous) BS equation, 
and also consistent with other indirect computation~\cite{Shuto:1989te} based on the ladder gauged Nambu-Jona-Lasinio model.

In fact, the PCDC (partially conserved dilatation current) relation evaluated in the ladder approximation 
does not allow a very light TD unless the TD gets decoupled with divergent decay constant $F_\phi$~\cite{Haba:2010hu,Hashimoto:2010nw}:  
The PCDC relation reads
\begin{equation} 
  F_\phi^2 M_\phi^2    =  -4\langle \theta^\mu_\mu \rangle 
  = \frac{\beta(\alpha)}{\alpha} \langle G_{\mu \nu}^2 \rangle
 \,, \label{PCDC}
   \end{equation}
   where $\theta_\mu^\mu$ is the trace of energy-momentum tensor. 
To this relation, the ladder estimate gives 
$ F_\phi^2 M_\phi^2  \simeq  3 \eta \,  m_F^4 $ (near the conformal window) with 
$\eta \simeq  \frac{N_{\rm TC} N_{\rm TF}}{2\pi^2}  ={\cal O} (1)$~\cite{Hashimoto:2010nw,Miransky:1989qc}.  
This simply implies that 
the limit $M_\phi/m_F \rightarrow 0 $,  where  the TD gets light compared with the weak scale 
$m_F  (={\cal O} (4 \pi F_\pi) )$,  can only be realized when $F_\phi/m_F  \rightarrow \infty$, i.e., a decoupled limit: 
\begin{equation} 
{\rm Ladder}: 
\qquad 
\frac{M_\phi}{F_\pi} \to 0  
\qquad 
\textrm{only when} 
\qquad 
\frac{F_\phi}{F_\pi} \to \infty
\,. \label{limit:lad}
\end{equation}

\subsection{Technidilaton Mass: Holographic Estimate}\label{holo-est}

As suggested earlier~\cite{Yamawaki:2007zz}, 
however, there is a potential problem in the ladder approximation about the mass of the TD.  
Actually, the ladder approximation totally ignores non-ladder dynamics, say, most notably the full gluonic dynamics.
Thus a possible way out would be to include fully {\it nonperturbative gluonic dynamics}. 
(Also a direct estimate of $F_\phi$ free from the ladder approximation and 
without invoking the PCDC (without referring to $M_\phi$)  
is necessary to give more implications of the TD at the LHC.)  
One such a possibility besides lattice simulations would be a holographic computation based on 
the gauge-gravity duality~\cite{Maldacena:1997re}.

We shall work on a recently proposed holographic model~\cite{Haba:2010hu}, which is  
based on deformation of 
a bottom-up approach for successful holographic-dual of QCD~\cite{DaRold:2005zs,Erlich:2005qh}. 
The model describes a five-dimensional gauge theory having 
$SU(N_{\rm TF})_L \times SU(N_{\rm TF})_R$ gauge symmetry, 
defined on the five-dimensional 
anti-de-Sitter space with the fifth direction being compactified on a finite interval.  
In addition to a bulk scalar dual to the technifermion bilinear operator $\bar{F} F$ with the anomalous 
dimension $\gamma_m =1$,  an extra bulk scalar field 
dual to technigluon condensate $\langle \alpha G_{\mu\nu}^2
\rangle$ is incorporated~\footnote{  
The extra bulk scalar dual to $\langle \alpha G_{\mu\nu}^2
\rangle$
 is actually necessary to properly reproduce 
the high energy behavior of 
vector/axial-vector current correlator including the gluon condensate term as predicted in 
the operator product expansion~\cite{Haba:2010hu}.}.

 Following the well-known holographic recipe, one can easily calculate 
 the scalar (or dilatation) current correlator and examine the lowest pole identified as the TD mass 
 to get a formula in a light TD limit:   
\begin{equation}
 {\rm Holography}: 
 \qquad 
\frac{M_\phi}{4\pi F_\pi} 
\simeq  \sqrt{ \frac{3}{N_{\rm TC}}} \frac{\sqrt{3}/2}{1+ G} 
\,, \qquad 
{\rm where} 
\qquad 
G \sim \frac{\langle \alpha G_{\mu\nu}^2   \rangle}{F_\pi^4}
\,. \label{holo:formula}
\end{equation} 
(The details can be found in the published paper~\cite{Matsuzaki:2012xx}.) 
This implies that, in contrast to the ladder approximation Eq.(\ref{mtd:lad}), 
 the TD can be as light as the 125 GeV boson by a large gluonic effect $G$:  
\begin{equation} 
  {\rm Holography}: 
\qquad 
M_\phi \simeq 125\, {\rm GeV} 
\qquad 
{\rm at} 
\qquad 
G \simeq 10
\, ,  
\end{equation} 
for the one-family model with $N_{\rm TC}=3$. 
It is also interesting to note that the gluonic effect in QCD is much smaller, 
$G|_{\rm QCD} \simeq 0.25$~\cite{Haba:2010hu}.

Similarly, one can calculate the TD decay constant $F_\phi$ by examining the pole residue of 
the dilatation current correlator to find a formula~\cite{Matsuzaki:2012xx}: 
\begin{equation} 
{\rm Holography}: 
\qquad 
 \frac{F_\phi}{F_\pi} \simeq \sqrt{ 2 N_{\rm TF}} 
 \,. \label{holo:formula:2}
\end{equation}
This and Eq.(\ref{holo:formula}) indicate that, in contrast to the ladder approximation Eq.(\ref{limit:lad}),  
 a very light/massless TD with the finite decay constant $F_\phi$ 
can be realized when $G$ goes to infinity: 
\begin{equation} 
{\rm Holography}: 
\qquad 
  \frac{M_\phi}{F_\pi} \to 0 \qquad 
  {\rm and} 
  \quad 
  \frac{F_\phi}{F_\pi} = {\rm constant} 
\,, 
  \quad 
  {\rm as} 
  \quad 
  G \to \infty 
  \,. 
\end{equation}

The large $G$ limit actually corresponds to a ``conformal" limit where $\beta(\alpha) \to 0$: 
Combining Eqs.(\ref{holo:formula}) and (\ref{holo:formula:2}) with the PCDC relation in Eq.(\ref{PCDC})
and calculating the technigluon condensate $\langle  \alpha  G_{\mu\nu}^2 \rangle$,  
one can evaluate the beta function as a function of $G$ to find that  
$  \beta(\alpha) \sim \frac{1}{G(1+G)^2} \to 0 $ as $G \to \infty$~\cite{Matsuzaki:2012xx}. 
Thus the holographic estimate suggests that 
a very light TD with the finite decay constant can be allowed due to the conformality tied with 
the nonperturbative gluonic effect $G$.

Though this holographic result is strong enough to indicate the presence of the 125 GeV TD, 
we would anticipate that lattice simulations will give a more conclusive answer.

\section{Low-Energy Effective Lagrangian}\label{eff-lag}

Below the technifermion mass scale $m_F = {\cal O}(4 \pi F_\pi) \sim 1$ TeV (See region III in Fig~.\ref{alpha-beta}), 
the WTC can be viewed as an effective technihadron model as in the QCD case. 
Such an effective model will be described by a low-energy effective Lagrangian based 
on nonlinear realization of both chiral and scale symmetries, involving the 125 GeV TD and the SM particles, 
which is most relevant for the LHC study~\footnote{
We focus only on the TD phenomenology with the SM particles and 
disregard technipions, some of which might participate in the low-energy model as discussed in a recent paper~\cite{Jia:2012kd}.   
}. 
The chiral/EW and scale invariant Lagrangian thus takes the form~\cite{Matsuzaki:2012vc}
\begin{equation}
{\cal L}_{\rm inv} = \frac{v_{\rm EW}^2}{4} \chi^2 {\rm tr}[D_\mu U^\dag D^\mu U] + {\cal L}_{\rm kin}(\chi) 
\,, \label{inv:L}
\end{equation}
where $\chi=e^{\phi/F_\phi}$ is a nonlinear base of the scale symmetry, which parametrizes the TD field $\phi$ with the decay constant $F_\phi$ 
and has the scale dimension 1; $D_\mu U= \partial_\mu U - i W_\mu U + i U B_\mu$ with the $SU(2)_W$ and $U(1)_Y$ gauge fields 
$W$ and $B$; ${\cal L}_{\rm kin}(\chi)$ denotes the scale invariant kinetic term of TD and $U$ the usual chiral field parameterizing 
the (eaten) Nambu-Goldstone boson fields $\pi$ as $U=e^{2 i \pi/v_{\rm EW}}$; 
The $|D_\mu U|^2$  term gives the TD couplings to massive weak bosons: 
\begin{equation} 
g_{\phi WW/ZZ} = \frac{2 m_{W/Z}^2}{F_\phi} 
 \,. \label{td:WW}
\end{equation}

As seen in Eqs.(\ref{NP:beta}) and (\ref{NP:SA}), 
the scale symmetry is actually broken explicitly as well as spontaneously  by 
dynamical  mass generation of technifermions, which   
 has to be respected also in the nonlinear realization~\cite{Matsuzaki:2012vc}. 
 Such explicit breaking effects arise in the TD Yukawa couplings to the SM fermions, 
which reflect underlying ETC-induced four-fermion terms, and couplings to QCD gluons and photons 
related to the scale anomaly in the SM gauge sector. 
In order to incorporate these effects into the scale-invariant Lagrangian, 
we introduce a spurion field $S$ having 
the scale dimension $1$ coupled to the SM fermions, digluon $gg$ and diphoton $\gamma\gamma$ 
in such a way that~\cite{Matsuzaki:2012vc} 
\begin{eqnarray} 
{\cal L}_{S} 
&=& 
- m_f \left( \left( \frac{\chi}{S} \right)^{2-\gamma_m} \cdot \chi \right) \bar{f} f 
\nonumber \\ 
&& 
+  \log \left( \frac{\chi}{S} \right) \left\{ 
\frac{\beta_F(g_s)}{2g_s} G_{\mu\nu}^2 
+ 
\frac{\beta_F(e)}{2e} F_{\mu\nu}^2 
\right\}  
\,, \label{Lag:S}
\end{eqnarray}
where $G_{\mu\nu}$ and $F_{\mu\nu}$ respectively denote the field strengths for QCD gluon and photon fields; 
$g_s$ and $e$ are the QCD gauge and electromagnetic couplings, respectively; 
$\beta_F$s are  the beta functions only including the technifermion 
loop contributions.

The TD Yukawa coupling to the SM $f$-fermion arises from the first line of Eq.(\ref{Lag:S}) as~\cite{Bando:1986bg} 
\begin{equation} 
 g_{\phi ff} =  \frac{(3-\gamma_m) m_f}{F_\phi} 
\,, \label{td:yukawa}
\end{equation}
along with scale dimension of technifermion bilinear operator 
$(3-\gamma_m)$. 
The anomalous dimension $\gamma_m \simeq 1$ in WTC, which is crucial to
obtain the realistic mass of the SM fermions of the first and the second generations without suffering from the 
flavor-changing neutral current problems.  
However it was known for long time that it is not enough for the mass of the third-generation 
SM $f$-fermions like $t, b, \tau$: 
A simplest resolution would be the strong ETC model~\cite{Miransky:1988gk} 
having much larger anomalous dimension $1<\gamma_m <2$ due to the strong effective four-fermion coupling
from the ETC dynamics 
in addition to the walking gauge coupling.  
Here we take $\gamma_m \simeq$ 2, i.e., $(3-\gamma_m) \simeq 1$, 
as in the strong ETC model 
for the third-generation SM $f$-fermions like $t, b, \tau$  
which are relevant to the current LHC data.

\subsection{Technidilaton Couplings to the SM particles}

We see from Eqs.(\ref{td:WW}) and (\ref{td:yukawa}) that 
the TD couplings to $W$ and $Z$ bosons and fermions are 
related to those of the SM Higgs by a simple scaling: 
\begin{eqnarray} 
  \frac{g_{\phi WW/ZZ}}{g_{ h_{\rm SM} WW/ZZ }} 
  &=& \frac{v_{\rm EW}}{F_\phi} 
  \,, \nonumber \\ 
  \frac{g_{\phi ff}}{g_{h_{\rm SM} ff}} 
  &=& \frac{v_{\rm EW}}{F_\phi} 
  \,, 
\qquad {\rm for} \quad f=t,b,\tau 
\,.  \label{scaling}
\end{eqnarray} 
In addition to the above scaling, 
the couplings to gluon and photon ($G_{\mu\nu}^2$ and $F_{\mu\nu}^2$ terms in Eq.(\ref{Lag:S}))  
involve the beta functions, $\beta_F(g_s)$ and $\beta_F(e)$, induced from 
$F$-technifermion loops~\cite{Matsuzaki:2011ie}.  
We shall employ the one-family model as a concrete WTC setting 
and evaluate the betas $\beta_F(g_s)$ and $\beta_F(e)$ at the one-loop level 
in the perturbation~\footnote{ This one-loop approximation can be justified 
in ladder estimate of the scale anomaly-related vertices $\phi$-$\gamma(g)$-$\gamma(g)$~\cite{Matsuzaki:2012vc}.}: 
\begin{eqnarray} 
\textrm{one-family model}: \qquad 
  \beta_F(g_s) = \frac{g_s^3}{(4\pi)^2} \frac{4}{3} N_{\rm TC}\,, \qquad  
 \beta_F(e) = \frac{e^3}{(4\pi)^2} \frac{16}{9} N_{\rm TC} 
\,.   \label{betas}
\end{eqnarray} 
We thus find the scaling from the SM Higgs for the couplings to 
$gg$ and $\gamma\gamma$, which can approximately be expressed 
 at around 125 GeV as  
\begin{equation}
\textrm{one-family model}: 
\qquad  
\begin{array}{ccc} 
\frac{g_{\phi gg}}{g_{h_{\rm SM} gg}} 
&\simeq &
\frac{v_{\rm EW}}{F_\phi} 
\cdot 
\left( 1 + 2 N_{\rm TC} \right)  
\\ 
\frac{g_{\phi \gamma\gamma}}{g_{h_{\rm SM} \gamma\gamma}} 
&\simeq & 
\frac{v_{\rm EW}}{F_\phi} 
\cdot 
 \left( 1 - \frac{32}{47} N_{\rm TC} \right)  
\end{array}
\,,  \label{g-dip-dig}
\end{equation} 
where in estimating the SM contributions  
we have incorporated only the top and $W$ boson loop contributions.

 One could actually derive these TD couplings directly from the Ward-Takahashi identities  
 for the dilatation symmetry (low-energy theorem)~\cite{Matsuzaki:2012vc}.

  As seen from Eqs.(\ref{scaling}) and (\ref{g-dip-dig}), 
once the ratio $v_{\rm EW}/F_\phi$ is fixed, 
the TD LHC phenomenological study can be made just by quoting 
the SM Higgs coupling properties: 
We may estimate $F_\phi$ based on the ladder approximation~\cite{Matsuzaki:2012vc}: 
 The TD decay constant $F_\phi$ can actually be related to the TD mass $M_\phi$ 
through the 
PCDC -- which is analogous to 
the PCAC (partially conserved axialvector current) 
relation for the QCD pion -- involving the vacuum energy density ${\cal E}_{\rm vac}$: 
\begin{equation} 
  F_\phi^2 M_\phi^2 = - 16{\cal E}_{\rm vac} 
\,. \label{Vac2}
\end{equation} 
 The vacuum energy density ${\cal E}_{\rm vac}$ is saturated by 
 the technigluon condensation induced from the technifermion condensation, 
 so is generically expressed as 
 \begin{equation} 
   {\cal E}_{\rm vac} 
   = - \kappa_V \left( \frac{N_{\rm TC} N_{\rm TF}}{8\pi^2} \right) m_F^4 
   \,, \label{Vac}
 \end{equation}
where $m_F$ denotes the dynamical technifermion mass and $N_{\rm TF}=2 N_{\rm D} + N_{\rm EW-singlet}$ 
with $N_D$ being the number of EW doublets formed by technifermions and  
the number of dummy technifermions, $N_{\rm  EW-singlet}$, which are singlet under the SM gauges. 
The overall coefficient $\kappa_V$ is determined once a straightforward nonperturbative calculation is made. 
The dynamical technifermion mass $m_F$ can, on the other hand, be related to 
the technipion decay constant $F_\pi$:  
\begin{equation} 
  F_\pi^2 = \kappa_F^2 \frac{N_{\rm TC}}{4 \pi^2} m_F^2 
\,,   \label{PS}
\end{equation}
with the overall coefficient $\kappa_F$ and the property of $N_{\rm TC}$ scaling taken into account. 
The values of $\kappa_V$ and $\kappa_F$ may be quoted from the latest result~\cite{Hashimoto:2010nw} on a 
ladder Schwinger-Dyson analysis for a modern version of 
WTC~\cite{Lane:1991qh,Appelquist:1996dq, Miransky:1996pd}: 
\begin{equation} 
 \kappa_V \simeq 0.7 \,, \qquad 
\kappa_F \simeq 1.4 
\,,  \label{kappas}
\end{equation} 
where $\kappa_F$ has been estimated based on the Pagels-Stokar formula~\cite{Pagels:1979hd}.  
In that case $N_{\rm TF}$ is fixed by the criticality condition 
for the walking regime as~\cite{Appelquist:1996dq} 
\begin{equation} 
 \frac{N_{\rm TF}}{4 N_{\rm TC}} \simeq 1 
 \,.
 \label{criticality}
\end{equation} 
The estimated values in Eqs.(\ref{kappas}) and (\ref{criticality}) 
are based on ladder approximation which are subject to certain uncertainties up to 30\%  
observed for the critical coupling and hadron spectrum in QCD~\cite{Appelquist:1988yc}.   
 We may include this 30\% uncertainty in estimation of each independent factor $\kappa_V$, $\kappa_F^2$   
and the criticality condition $N_{\rm TF}/(4N_{\rm TC})$. 
  Putting these all together, we thus estimate $v_{\rm EW}/F_\phi$ as  
\begin{eqnarray} 
\textrm{one-family model}: 
\qquad 
\frac{v_{\rm EW}}{F_\phi}\Bigg|_{\rm Ladder}
&\simeq& 
(0.1 - 0.3) \times  \left( \frac{N_D}{4} \right) \left( \frac{M_\phi}{125\,{\rm GeV}} \right) 
\,, \label{vals}
\end{eqnarray}

Alternatively, using the holographic estimate in Eq.(\ref{holo:formula:2}) 
and incorporating a typical $\sim $30\% correction into the holography 
coming from the next-to-leading order terms in 
$1/N_{\rm TC}$ expansion,  
we may estimate the TD decay constant $F_\phi$ to get~\cite{Matsuzaki:2012xx} 
\begin{equation} 
\textrm{one-family model}: 
\qquad  
\frac{v_{\rm EW}}{F_\phi}\Bigg|^{+1/N_{\rm TC}}_{\rm holo} 
\simeq 0.2 -0.4 
\, . \label{holo:prediction}
\end{equation} 
which is coincidentally consistent with the ladder estimate in Eq.(\ref{vals}). 
Note that the two calculations are quite 
different qualitatively in a sense that 
the ladder calculation has no massless TD limit, while 
the holographic model including the nonperturbative 
gluonic dynamics does. Nevertheless, such a numerical coincidence 
may suggest that 
both models are reflecting some reality through similar dynamical effects 
for the particular mass region of the 125 GeV TD.

\subsection{Technidilaton Mass Stability} 

Before proceeding to the LHC phenomenology, 
we shall briefly remark on  stability of the light TD mass against radiative corrections.  
As a pseudo Nambu-Goldstone boson of scale invariance, the quadratic divergence is suppressed by the 
scale invariance for the walking regime $m_F < \mu <\Lambda_{\rm TC}(\sim \Lambda_{\rm ETC})$ (region II in Fig.~\ref{alpha-beta}). 
The scale symmetry breaking in the ultraviolet region $\mu>\Lambda_{\rm TC}$ (region I in Fig.~\ref{alpha-beta})  has no problem thanks to 
the naturalness as usual like in the QCD and the QCD-scale-up TC where the  theory has  only logarithmic divergences. 
Only possible source of the scale symmetry violation is from an effective theory for $\mu < m_F$ (region III in Fig.~\ref{alpha-beta}).

Below $\mu=m_F$ 
the dominant corrections to the TD mass $M_\phi$ come from the SM top quark loop~\footnote{
 Actually, a TD potential term involving the usual $\lambda \phi^4$ coupling could yield 
 a similar correction to $M_\phi$ at one-loop level. This contribution, however, turns out to be 
negligible in magnitude compared to the top-loop correction in Eq.(\ref{top-corr}), due to the high suppression factor 
$(M_\phi/(4\pi F_\phi))^2$~\cite{Matsuzaki:2012vc}. }, 
which can be estimated from the effective Lagrangian in Eqs.(\ref{inv:L}) and (\ref{Lag:S}) as 
\begin{equation} 
 \delta M_\phi^2 \simeq 
 - \frac{3}{4\pi^2} \frac{m_t^2}{F_\phi^2} \, m_F^2 
 \,, \label{top-corr}
\end{equation} 
where the cutoff has been set by $m_F$. 
Due to the large TD decay constant $F_\phi \sim 1 {\rm TeV}(\simeq m_F={\cal O}(4\pi F_\pi))$ (see Eq.(\ref{holo:prediction})), 
this correction only gives a tiny shift to the 125 GeV TD mass (Note numerically $m_t^2 \simeq 2 M_\phi^2$): 
\begin{equation} 
  \Bigg| \frac{\delta M_\phi}{M_\phi(125{\rm GeV})} \Bigg| 
\simeq  \frac{3}{4\pi^2} \frac{m_F^2}{F_\phi^2} 
\simeq {\cal O}(10^{-2}-10^{-1}) 
\,. 
\end{equation}
Thus the 125 GeV TD mass is fully stable against the radiative corrections, in contrast to 
the unnatural SM Higgs case.

\section{125 GeV Technidilaton Signal at the LHC} \label{signal}

We now discuss the 125 GeV TD phenomenology at the LHC based on the coupling properties 
derived in the previous section. 
As is clear from Eqs.(\ref{scaling}) and (\ref{holo:prediction}), 
the TD couplings to the SM particles are simply suppressed by the large decay constant $F_\phi$
compared to the SM Higgs case. 
As seen from Eq.(\ref{g-dip-dig}), however, the gluon-gluon fusion (ggF) and diphoton couplings can be 
enhanced due to extra contributions from technifermions in the one-family model 
which compensate the overall suppression by $F_\phi$. 
Thus the ggF process becomes mostly dominant in the TD LHC production, 
while other production processes such as vector boson fusion (VBF) and vector boson associate production (VH) are 
suppressed simply by the large TD decay constant $F_\phi$. 
Similarly, the branching fraction is almost governed by the decay to digluon which becomes 
comparable with $b\bar{b}$ channel for the SM Higgs, while other channels such as 
the decay to $WW^*, ZZ^*$ and fermion pairs are suppressed. (The TD total width at 125 GeV is comparable with the SM Higgs one.)
More details on the production and decay properties can be found in the published papers~\cite{Matsuzaki:2012vc}.

In order to make a direct comparison with the current LHC data~\cite{Aad:2012tfa,ATLAS-CONF-2012-168,ATLAS-CONF-2012-160},  
we estimate the TD signal strengths normalized to the SM Higgs cross section, 
\begin{equation} 
\mu_X = \frac{\sigma_\phi(pp \to \phi) \times {\rm BR}(\phi \to X)}{
\sigma_{h_{\rm SM}}(pp \to h_{\rm SM}) \times {\rm BR}(h_{\rm SM} \to X)}
\, . 
\end{equation} 
Taking the one-family model with $N_{\rm TC}=4$ and $v_{\rm EW}/F_\phi=0.2$ as a benchmark, 
in Table~\ref{signal-strength} we summarize the predicted signal strengths at 125 GeV  
for each category relevant to the current Higgs search at 8TeV LHC. 
  As for VBF-tag categories, we have taken into account 
  about 30\% contamination from the ggF process as in the case of the SM Higgs.

\begin{table}
 \tbl{The predicted 8TeV LHC signal strengths of 125 GeV TD in the one-family model with $N_{\rm TC}=4$ 
and $v_{\rm EW}/F_\phi = 0.2$ fixed. }
{
\begin{tabular}{|c||c|c|c|}
\hline 
\hspace{12pt} channel \hspace{12pt} &\hspace{12pt}  ggF-tag \hspace{12pt}  & \hspace{12pt}  VBF-tag \hspace{12pt}  & \hspace{12pt}  VH-tag \hspace{12pt}  \\ 
\hline \hline 
$\gamma\gamma$ & 1.4 & 0.4 & 0.02 \\ 
\hline 
$ZZ^*$ & 1.0 & 0.3 & 0.01 \\ 
\hline 
$WW^*$ & 1.0 & 0.3 & 0.01 \\ 
\hline 
$\tau^+\tau^-$ & 1.0 & 0.3 & 0.01 \\ 
\hline 
$b\bar{b}$ & --- & --- & 0.01 \\ 
\hline 
\end{tabular}
} 
\label{signal-strength}
\end{table}

From Table~\ref{signal-strength} we see that 
the diphoton signal strength for the ggF-tag category is slightly enhanced 
and can explain the currently reported excess at the LHC~\cite{Aad:2012tfa,ATLAS-CONF-2012-168,ATLAS-CONF-2012-160}, while     
other channels in the ggF-tag categories are almost SM Higgs-like, not distinguishable 
from those for the SM Higgs. 
The characteristic feature of the 125 GeV TD in the one-family model is seen at 
the VBF- and VH-tag categories, which are significantly suppressed compared to the 
SM Higgs case, simply due to the overall suppression of the couplings by the large decay constant $F_\phi$. 
This will be tested in the future, 
though the presently observed significance on such categories are 
still lower than those for the ggF-tag category.

\begin{figure}[t]
\includegraphics[scale=0.7]{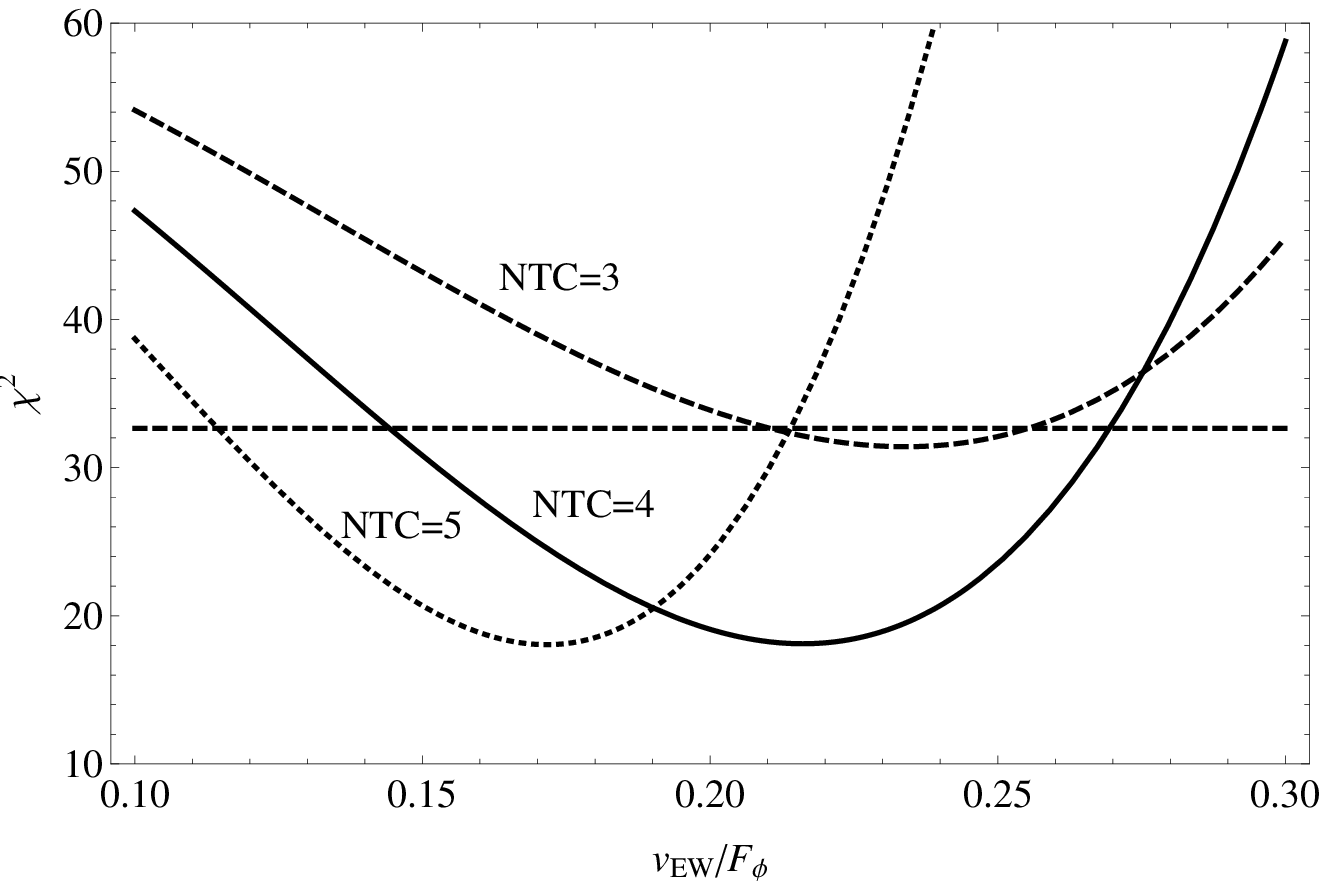}
\includegraphics[scale=0.55]{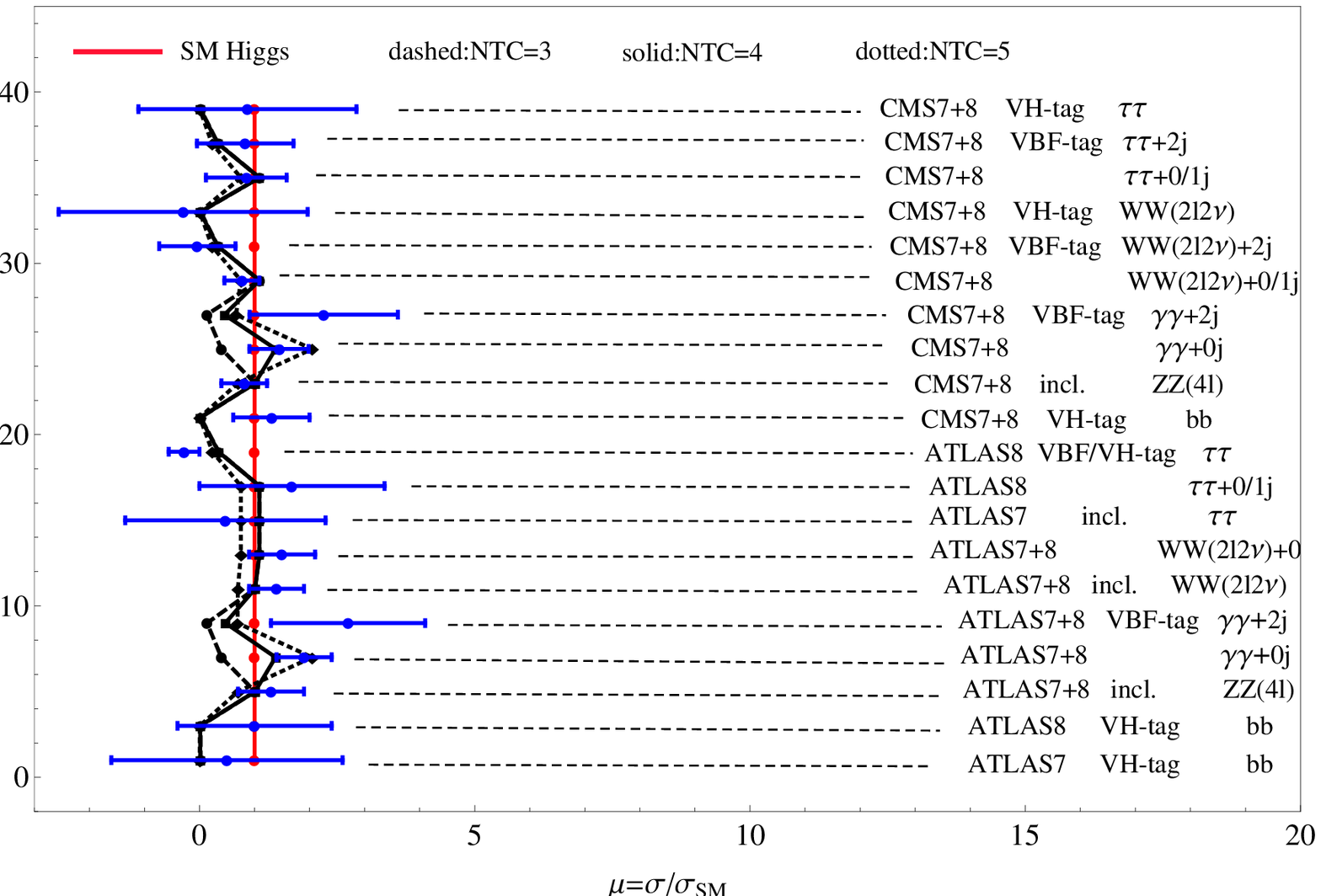}
\caption{Top panel: $\chi^2$ as a function of $v_{\rm EW}/F_\phi$ for the 125 GeV TD in the one-family model with 
$N_{\rm TC}=3,4,5$. The dotted straight line corresponds to the SM Higgs case. 
Bottom panel: the predicted signal strengths in comparison with 
the current LHC data~\cite{Aad:2012tfa,ATLAS-CONF-2012-168,ATLAS-CONF-2012-160}. }
\label{fit-afterHCP}
\end{figure}

Finally, we shall test the goodness-of-fit of the TD with use of the current LHC data~\cite{Aad:2012tfa,ATLAS-CONF-2012-168,ATLAS-CONF-2012-160}, 
based on the $\chi^2$ function: 
\begin{equation} 
  \chi^2 = \sum_{i \in {\rm events}} \left( \frac{\mu_i - \mu_i^{\rm exp} }{\sigma_i} \right)^2  
  \,, 
\end{equation}
where $\mu_i^{\rm exp}$ denote the best-fit strengths for each channel reported 
and $\sigma_i$ the corresponding one sigma errors. 
Taking $(v_{\rm EW}/F_\phi)$ as a free parameter so as to satisfy the theoretically expected range in Eq.(\ref{vals}) (or Eq.(\ref{holo:prediction})), 
 in Fig.~\ref{fit-afterHCP} we plot the $\chi^2$ function for the 125 GeV TD in 
the one-family model with $N_{\rm TC}=3,4,5$. 
In particular, the best-fit values for $N_{\rm TC}=4,5$
are as follows: 
\begin{eqnarray} 
\begin{array}{c|cc} 
\hline 
 N_{\rm TC}  & \hspace{15pt}  (v_{\rm EW}/F_\phi)_{\rm best} \hspace{15pt}  
& \hspace{15pt}  \chi^2_{\rm min}/{\rm d.o.f}  \hspace{15pt}  \\    
\hline 
4 & 0.22 & 18/19 \simeq 0.95  \\ 
5 & 0.17 & 18/19 \simeq 0.95  \\ 
\hline 
\end{array}
\,, 
\end{eqnarray}
which are compared with the SM Higgs case, $\chi^2|_{\rm SM} \simeq  33/20 \simeq 1.6$, 
implying that the TD is more favorable than the SM Higgs $(\mu_i=1)$. 
 This nice goodness of fit is due to the significant enhancement in the diphoton channel 
coming from the sector beyond the SM (technifermions): 
The technifermion loop contributions as in Eq.(\ref{g-dip-dig}) 
become large enough to compensate  the smallness of the overall $(v_{\rm EW}/F_\phi)$ in Eq.(\ref{vals}).

\section{Conclusion}\label{conc}

The TD is the characteristic light composite scalar in WTC. 
The mass can be 125 GeV, protected by the approximate scale invariance. 
The TD mass is also stable against the SM radiative corrections, 
where the large TD decay constant plays an important role. 
The couplings to the SM particles take essentially the same form as those for the SM Higgs, 
except couplings to diphoton and digluon which involve a high model-dependence. 
The 125 GeV TD in the one-family model gives the LHC signal consistent with the LHC data, 
notably can explain the currently reported diphoton excess. 
More precise measurements in VBF and VH categories will draw a definite conclusion that the TD is favored, or not.  
\\

{\it Note added}: 

After the SCGT 12 workshop, 
the ATLAS and CMS have reported updated data in some channels for 
the Higgs search involving $WW^*, ZZ^*, \tau^+\tau^-$ and diphoton channels~\cite{update,Moriond-Aspen}. 
However, the signals in VBF- and VH-tag categories have not yet clearly been measured,  
which would actually be the most characteristic for the 125 GeV TD 
in the one-family model: 
The most significant channels with the significance over $\sim 3 - 4$ sigma 
are still only in the ggF-tag categories, 
as far as the amount of observed excess over the background is concerned. 
See Table~\ref{signal-strength:added}. 
Thus the 125 GeV TD in the one-family model is still consistent with the latest data. 
Further update on measurements in the VBF- and VH-tag categories is needed to more definitely say 
that is the TD, or not.

\begin{table}
 \tbl{The predicted 8TeV LHC signal strengths of 125 GeV TD in the one-family model with $N_{\rm TC}=4$ 
and $v_{\rm EW}/F_\phi = 0.2$ fixed, in comparison with the latest LHC data as of {\it Moriond 2013 EW and QCD sessions}   
(or {\it Aspen 2013-Higgs Quo Vadis}) in 2013~\cite{update,Moriond-Aspen}. 
The sizes of the observed significance for signal over background have also been displayed. } 
{
\begin{tabular}{|c||c|cc|cc|}
\hline 
\hspace{5pt} category \hspace{5pt} &
\hspace{5pt} TD signal strength \hspace{5pt} &
\hspace{5pt}  ATLAS  \hspace{5pt} &\hspace{5pt}  significance  \hspace{5pt}  &
 \hspace{5pt}  CMS  \hspace{5pt} & \hspace{5pt} significance \hspace{5pt}  \\ 
\hline \hline 
ggF-tag & & &  \\ 
\hline 
$\mu_{\gamma\gamma}^{\rm ggF}$ & 1.4 & $1.6 \pm 0.4$ & $\sim 6 \sigma$ & $0.5 \pm 0.5$ & $\sim 3 \sigma$ \\ 
\hline 
$\mu_{WW^*}^{\rm ggF}$ & 1.0 & $0.8 \pm 0.4$ & $\sim 4 \sigma$  & $0.8\pm 0.2$  & $\sim 4 \sigma$ \\ 
\hline 
$\mu_{ZZ^*}^{\rm ggF}$ & 1.0 & $1.8^{+0.8}_{-0.5}$ & $\sim 6 \sigma$ & $0.9^{+0.5}_{-0.4}$ & $\sim 6 \sigma$ \\ 
\hline 
$\mu_{\tau^+ \tau^-}^{\rm ggF}$ & 1.0 & $2.1^{+4.0}_{-3.0}$ & $< 2 \sigma$  & $0.7 \pm 0.5$ & $\sim 3 \sigma$  \\ 
\hline  
VBF-tag & & &  \\ 
\hline 
$\mu_{\gamma\gamma}^{\rm VBF}$ & 0.4 & $1.7 \pm 0.9$ & $< 2 \sigma$ with VH & $1.5^{+1.5}_{-1.1}$ &$ < 2 \sigma$ with VH \\ 
\hline 
$\mu_{WW^*}^{\rm VBF}$ & 0.3 & $1.7 \pm 0.8$ & $< 2 \sigma$ & $0.04^{+0.77}_{-0.57}$ & $< 2 \sigma$ \\ 
\hline 
$\mu_{ZZ^*}^{\rm VBF}$ & 0.3 & $1.2^{+3.8}_{-1.4}$ & $< 2 \sigma$ with VH & $1.2\pm 5.6$ & $< 2 \sigma$ \\ 
\hline 
$\mu_{\tau^+ \tau^-}^{\rm VBF}$ & 0.3 & $-0.4^{+3.0}_{-2.0}$ & $< 2 \sigma$ with VH & $1.4 \pm 0.6$ & $< 2 \sigma$ \\ 
\hline 
VH-tag & & &  \\ 
\hline 
$\mu_{\gamma\gamma}^{\rm VH}$ & 0.02 & $1.8^{+1.5}_{-1.3}$ & $< 2 \sigma$ with VBF & $1.5^{+1.5}_{-1.1}$ & $< 2 \sigma$ with VBF \\ 
\hline 
$\mu_{WW^*}^{\rm VH}$ & 0.01 & --- & --- & $-0.3^{+2.3}_{-2.0}$ & $< 2 \sigma$ \\ 
\hline 
$\mu_{ZZ^*}^{\rm VH}$ & 0.01 & $1.2^{+3.8}_{-1.4}$ & $< 2 \sigma$ with VBF & --- & --- \\ 
\hline 
$\mu_{\tau^+ \tau^-}^{\rm VH}$ & 0.01 & $-0.4^{+3.0}_{-2.0}$ & $< 2 \sigma$ with VBF & $0.8^{+1.5}_{-1.4}$ & $< 2 \sigma$ \\ 
\hline 
$\mu_{b\bar{b}}^{\rm VH}$ & 0.01 & $-0.4 \pm 1.0$ & $< 2 \sigma$ & $1.3^{+0.7}_{-0.6}$ & $< 2 \sigma$ \\ 
\hline 
\end{tabular}
} 
\label{signal-strength:added}
\end{table}

\section*{Acknowledgments}

I would like to thank K.~Yamawaki for collaborations for a couple of recent papers on the technidilaton phenomenology. 
This work was supported by 
the JSPS Grant-in-Aid for Scientific Research (S) \#22224003. 


\end{document}